\documentclass[prl,preprint,showpacs,byrevtex,superscriptaddress]{revtex4}
\begin{document}
\title {Shape anisotropy and Voids }
\author{Gauri R. Pradhan}
\email{gau@physics.unipune.ernet.in}
\affiliation{Department of Physics, University of Pune, Pune 411 007,
  India}
\author{Sagar A. Pandit}
\email{sagar@prl.ernet.in}
\affiliation{ Physical Research Laboratory, Ahmedabad 380 009, India}
\author{Anil D. Gangal}
\affiliation{Department of Physics, University of Pune, Pune 411 007,
  India}
\author{V. Sitaramam}
\email{sitaram@unipune.ernet.in}
\affiliation{Department of Biotechnology, University of Pune, Pune 411 
  007, India} 
\begin{abstract}

Numerical simulations  on  a 2-dimensional model system showed that
voids are induced primarily due to shape anisotropy in binary mixtures
of interacting disks. The results of such a simple model 
account for the key features seen in a variety of flux
experiments using liposomes and biological membranes~\cite{Sita}.  
\end{abstract}
\maketitle

A variety of lipid molecules contribute to the structure and barrier 
function of biological membranes. Part of each
molecule is hydrophilic (head) and part hydrophobic (tail). This
amphipathic nature, in the presence of water, leads to a bilayer
structure in which hydrophilic head groups 
have maximum contact with water and hydrophobic tails have minimum
contact with water. The process of membrane formation is one
of minimizing the free energy and maximizing the stability of the
structure~\cite{Stein}.  

        A major question has been whether the bilayer is to be viewed
as an isotropic homogeneous phase or as a heterogeneous
phase~\cite{Lee}. If osmotic contraction
of the bilayer vesicles leads to an altered hydraulic conductivity
(water flux coefficient), one obviously favors a heterogeneous
membrane model. Otherwise a homogeneous, isotropic phase model would
be adequate, obviating the need to look for a fine structure within
the bilayer. Using the erythrocyte  as an experimental system ( in
which the area of the biconcave cell does not change when it is
osmotically expanded to a spherical shape), it was concluded that
hydraulic conductivity was stretch independent, i.e., in support of
the isotropic model~\cite{Sita}. An alternative way to assess hydraulic
conductivity  is to use hydrogen peroxide as an analog of water: since
many experimental systems have catalase (an enzyme that degrades
hydrogen peroxide to molecular oxygen and water) within the
vesicle/cell, an assay of this occluded catalase directly permits one
to measure the conductivity to exogenous hydrogen 
peroxide. Under equilibrium conditions of assay,
the rate of degradation would be same as the rate of permeation of the
peroxide into the vesicle. Thus, one can directly assess the stretch
sensitivity of the membrane by osmotic titrations with osmolites, using
non-electrolytes like hydrogen peroxide as probes of flux. In the course of
these experiments it was found that~\cite{Sita}: (i) among all the
lipid combinations tested, the phosphatidylcholine (PC) vesicles and
intact erythrocytes, both, did not show a decrease in occluded catalase
activity on osmotic compression of the membrane (ii) on the other
hand, all other membrane systems, such as 
peroxisomes, E.coli, macrophages showed stretch (osmotic) sensitivity
(iii) so did liposomes made from these cells and organelles (iv)
further, when binary mixtures were investigated, only cardiolipin and
cerebrosides when added to PC (5 to 10$\%$ of PC) conferred stretch
sensitivity in liposomes (v) these binary mixtures also exhibited enhanced
activation volume(osmotic sensitivity) and diminished activation energy
for hydrogen peroxide flux (vi) further, glucose was readily
permeable across these membranes of binary mixtures (vii)  addition of
cholesterol, which is abundantly found in erythrocytes, inhibited the stretch
sensitivity to peroxide permeation (viii) evidence was also seen that
this diffusion increases with decrease in temperature, i.e., the
process has a negative temperature coefficient.

These studies on biological membranes and liposomes, in which
composition as well as dynamics considerably vary, prompted us to
question as to what constitutes a minimal description to account for
variable permeability induced by doping across a liposomal membrane.
For instance, cardiolipin enhanced permeation of hydrogen
peroxide and molecules as large as {\it glucose}~\cite{Sita}. Though possible descriptors 
could be many, (composition, structure, dynamics in terms of inter
and intra molecular potentials), we adopted a bare-bone  
approach to resolve this complex issue to arrive at a minimal
description adequate to account for the observations. 

        Structural changes in the membrane are best identified by
non-interactive molecules and therefore leaks across
bilayers are commonly studied using non-electrolytes~\cite{Stein}. The diagnostic
for non-specific permeation is size dependence such that, these
hydrated solutes intercalate, penetrate and navigate through such
interstices, spaces or voids, stochastically or in files to reach the
other side of the membrane~\cite{Lee}. In order to capture the diverse features
in a parsimonious manner, we restrict to a two dimensional
cross-section of a three dimensional system. Such a restriction is reasonable since 
the probe particle permeating across the membrane  at any instant of time
experiences the effective 
cross section rather than the three dimensional obstruction. The  
permeation across the membrane depends primarily on the availability
of free space or voids. Thus the problem reduces to the study of 
packing of 2-dimensional objects at the first instance. Then one needs
to determine which factor(s) determine the appearance and 
size-distribution of voids in such a 2-dimensional system. 

The configuration space of this model system (membrane) is a
2-dimensional box with 
toroidal boundary conditions. The constituents of this two dimensional box
are the circular disks (and/or the rigid combinations of the circular
disks as dopants) of unit radii. A circular disk simply represents
the hard core scattering cross  section, seen by the passing
particle (a non-electrolyte which acts as a probe), across the
thickness spanned by two lipid molecules, viewed somewhat as
cylinders. A typical dopant is two or more circular disks rigidly joined
in a
prespecified geometry. These constituents are 
identified by the position coordinates of their centers, the angle
made by the major axes with the side of the box ( in case of dopants)
and the radii of the circular disks.  
It is reasonable that the disks (which represent molecules,
with long range attractive interaction and hard core repulsion near
center, contained within a structure) 
interact pair wise via {\it Lennard-Jones} potential (a measure of the
interaction energy), which has the form  
\begin{eqnarray}
%\label{LJ}
V_{LJ}(r_{ij}) = 4 \epsilon \sum_{i=1}^N \sum_{j = i+1}^N \Big( ({\sigma
\over r_{ij}} )^{12} - ( { \sigma \over r_{ij}})^6 \Big) \nonumber
\end{eqnarray}
where, $r_{ij}$ is the distance between the centers of the $i^{\rm
th}$ and $j^{\rm th}$ disks, $\sigma$ determines the range of hard
core part in the potential and the $\epsilon$ signifies the depth of
attractive part. 
While studying the binary mixtures, we consider different {\it shape
  anisotropic} combinations ({\it impurities} or {\it dopants}) of
$\kappa$ number of circular disks. We treat these combinations as one
unit.  
e.g. {\it rod$_n$} denotes a single dopant made of $n$ unit circular disks rigidly
joined  one after another  in a straight line. 
The impurities interact with constituent circular disks via potential,
\begin{eqnarray}
V(r_{ij}) = \sum_{\alpha = 1}^\kappa V_{LJ} (r_{i_\alpha j}) \nonumber
\end{eqnarray}
where, $r_{i_\alpha j}$ is the distance between the centers of
$\alpha^{\rm th}$ disk in $i^{\rm th}$ impurity and the $j^{\rm th}$
circular disk, and among themselves interact via
\begin{eqnarray}
V(r_{ij}) = \sum_{\alpha = 1}^{\kappa_1} \sum_{\beta = 1}^{\kappa_2}
V_{LJ}(r_{i_\alpha j_\beta})\nonumber
\end{eqnarray}
where, $r_{i_\alpha j_\beta}$ is the distance between the centers of
$\alpha^{\rm th}$ disk in $i^{\rm th}$ impurity and the $\beta^{\rm
th}$ disk in $j^{\rm th}$ impurity.

An $r$-void  is defined as a closed area in a membrane
devoid of disks or impurities, and sufficient to accommodate a
circular disk of radius $r$~\cite{Gauri}. Clearly, larger voids also
accommodate smaller probes, i.e., an $r$-void is 
also an $r^\prime$-void if $r^\prime < r$. Similarly, the
voids for the particle of size zero are the voids defined in the
conventional sense, i.e., a measure of the net space unoccupied by the
disks.

Equilibrium configurations of the  model system are obtained by a Monte
Carlo method (using the Metropolis algorithm) starting with a random
placement of the disks~\cite{Gauri,Binder}\footnote{The equilibrium
configurations thus obtained are further confirmed by simulated
  annealing~\cite{NR}.}. The box was filled with disks such that they
occupy 70\% area of the box,i.e.,loosely packed to facilitate the
formation of voids.  The temperature parameter, $T$, was so chosen
that the quantity $k_B T < 4 \epsilon$. This ensured an approximate
hexagonal arrangement of the disks and  the presence of very few large voids in the
absence of dopants. (Fig. 1a; far too less r-voids for $r \geq 0.5$\footnote {It
may be recalled that glucose offers approximately half the radius of
the PC cross section, yielding a relevant definition for a larger void of
interest.}). Similar numerical simulations are performed 
on the model system with dopants. The number of dopants is chosen to
be 10\%~\cite{Sita} of the number of circular disks with a
constraint that the total occupied area of the box is still 70\% as
the focus was on the redistribution of void sizes. Fig. 1b illustrates
the formation of larger voids in the vicinity of the rod$_2$ impurities. 

The variation in the number of $r$-voids  as a function of the
size of the permeating particle(using the digitization algorithm
described in~\cite{Gauri}) is shown in Fig. 2. When only circular 
disks are present (dotted curve), hardy any large $r$-voids are seen. When
mixed with the anisotropic impurities, say, rod$_2$, a distinct
increase in the number of large $r$-voids  is seen with appropriate
redistribution of smaller $r$-voids (solid curve). This result
is consistent with the unexpected permeation of large
molecules such as glucose through the doped membrane, observed
experimentally~\cite{Sita}.The difference curve showed the formation
of a significant extent ($\approx$ 30\%) of $r$-voids of size $0.5$
and above.   

Is the induction of large voids due to the anisotropy
in potential of the impurities and, should the large voids form around
the rods, the centers of anisotropy? Firstly, we carried
out simulations with large circular disks in place of rod$_2$ as
impurities. The radius of large disk was chosen in such a way that the
area occupied by each of the large disk is same as that of a
rod$_2$.  Fig. 3 shows the result of such simulations. The curve (a)
in Fig. 3 represents the difference curve of $r$-void distribution of
pure membrane and that of membrane doped with rod$_2$ impurity.
 The curve (b) represents the same when the membrane is doped with
large circular disks. It can be clearly seen 
that the number of larger $r$-voids is always less in the latter case,
thus confirming the role of shape anisotropy in the induction
of large $r$-voids.
Further, simulations are carried out with rod$_2$ of smaller size and
rod$_4$ type impurities. The curves (c), (d) in Fig. 3 respectively
shows the corresponding difference curves. It shows an interesting
feature that the peak of the difference curve shifts with the change
in the type of anisotropy. This suggests a possible way of
constructing membrane with selective permeability properties. 
The simulation regime adopted here, limited the exploration of ternary
mixtures in yielding statistically significant results on
transport. However, by using rod$_2$ of 0.5 size (Fig.3, curve (c)) (an oval
approximation of the small dopant cholesterol), we could
demonstrate a shift in void sizes to left in binary mixtures,
consistent with our experimental results in ternary mixtures with
cholesterol~\cite{Sita}. 

Further, we considered rod$_n$ type
impurities. Fig. 4 represents the relation between the length
dimension of rod$_n$ and the number of $r$-voids (for r=0.55). 
The anisotropy in the potential of rod$_n$ increases with $n$, such
that the number of large $r$-voids should increase
with $n$. Fig. 4 indeed shows a jump when the rod$_2$ impurities are
added and afterwards, it shows a slow and almost linear increase with
increase in $n$.

Since dopants induced voids, their influence is most likely to be seen 
in their own vicinity, enhancing the ``local transport''. As the
dopants exhibit different potential in different directions, certain
positions of the constituents are preferred from the   point of view
of energy minimization, eventually giving rise to voids in the vicinity of impurities. 
To verify this, we calculate the {\it local permeation probability}
for particle of size $r$, which is a ratio of the area of $r$-voids
and the area of the local neighborhood. Fig. 5 shows  the local
permeation probability around ten randomly chosen impurities and
ten randomly chosen circular disks. The higher local permeation
probability is seen to be associated with rod$_2$.

The model is realistic in that, one can compute elastic properties (surface tension
like attribute) by stretching the membrane from one side and computing
the energy change, which yielded a change of $\approx$ 13 dyn/cm which is of
the same order as the observed surface tension and changes there of in
bilayers~\cite{Jan}. The model is limited in relation to an investigation of
temperature effects which requires incorporation of multiple time
scales. The model is really general because it not restricted by the
parameter space of the components and therefore it is extendible  to a
variety of phenomena including transport in weakly
bound granular media, voids in polymers, modeling of zeolites which
may act like a sponge absorbing only desired species.

In summary, we proposed a two dimensional computational model system
comprising a mixture of objects interacting via Lennard-Jones
potential to explain anomalous  permeation seen in bilayers. The
significance of these observations on permeation, in what is
essentially a granular medium (with long range attraction as an added
feature), relates to development of large voids  seeded by
impurities. Unlike shape anisotropy, the change in composition (via a change 
in $\sigma$), lateral pressure/density, and presumably temperature (though
dynamic simulations would need to be done) would all simply produce
voids within the bounds of a hexagonal array. Even among various
dopants in shapes, X, L, Y, Z, T symmetrical or otherwise, all other factors being
equal, size per size, rod$_n$ produce the largest voids~\cite{Gauri}. 

Thus the largest $r$-voids induced by anisotropy yield biological
observations of interest, without manifesting as ordered
geometric structures. In this sense these $r$-voids differ from the conventional
results obtained in studies on granular media (which also
do not usually incorporate long range interactions)~\cite{Granular}. 
It is increasingly becoming clear that  voids play a pivotal role in
relating the dynamics of biopolymers to specific functional states~\cite{Raj}.

We thank C. N. Madhavarao for discussions.  The author (GRP) is
grateful to CSIR (India) for fellowship and (ADG) is grateful to DBT
(India) for research grant.

\newpage
\centerline{\bf FIGURE CAPTIONS}
\begin{description}
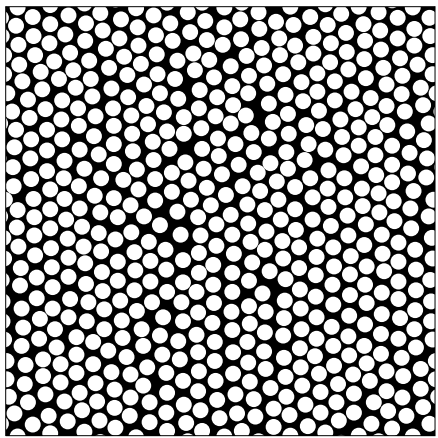
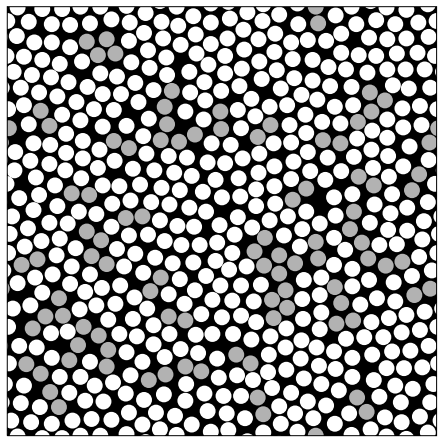
\item[{\bf Fig. 1}] Typical equilibrium configurations for interacting disks.
The parameter, $\sigma$ in Lennard-Jones potential is 2 units. The size of the
box is 50$\times$50 (a)  A pure membrane with 556 unit circular disks. 
(b) A doped membrane with 464 circular disks and 46 rod$_2$(shown in
gray).

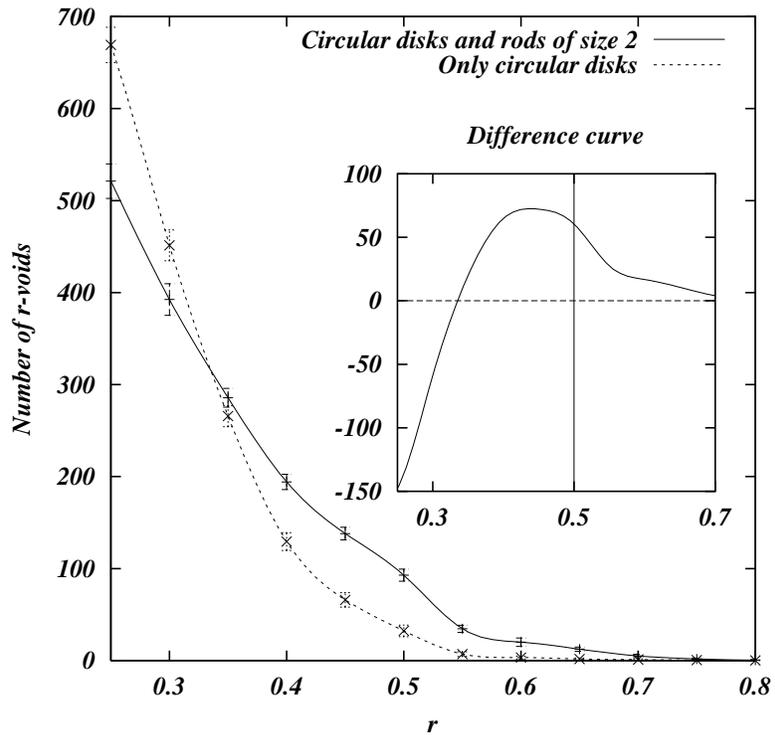
\item[{\bf Fig. 2}] Distribution of $r$-voids in two different
configurations. The main graph shows the number of $r$-voids as a
function of the relative size ($r$) of the probe particle. The vertical
bars represent the error margins at the corresponding points. 
the dotted curve gives the distribution in pure membrane while the
solid curve shows the same in a membrane doped with rod$_2$
(10:1). Difference curve clearly demonstrates the presence of large voids in
the doped membrane. 

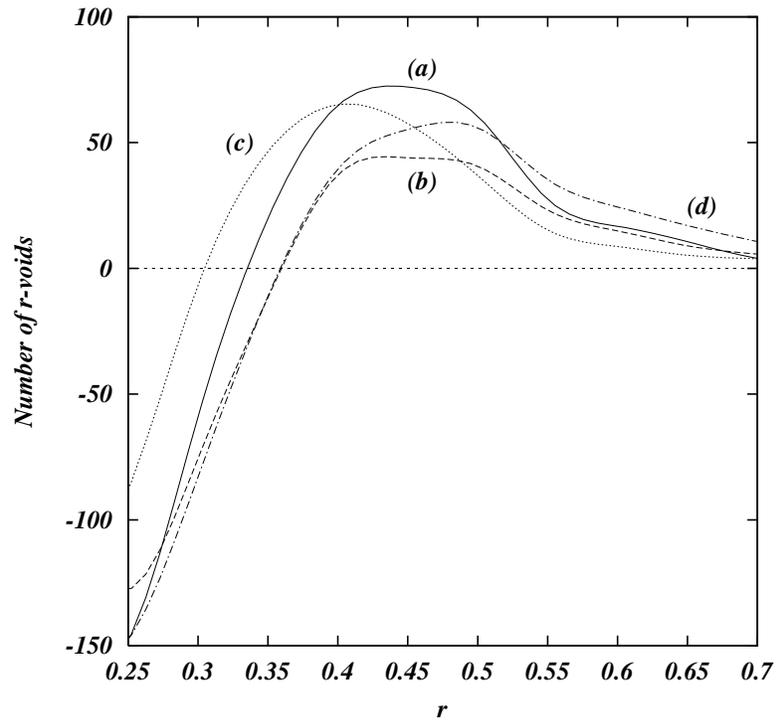
\item[{\bf Fig. 3}] Difference curves of distribution of $r$-voids. Curves are
obtained by treating the void distribution for pure membrane as the base. 
Difference curves for membranes(a) doped with rod$_2$
(b) doped with large circular disks (of radii $\sqrt{2}$) which occupy 
the same area as that of rod$_2$. These induce smaller number of
large voids as compared to rod$_2$.
(c) doped with small rods. Irrespective of their small size, they
induce large voids, but the peak shifts towards left.
(d) doped with rod$_4$. Significantly large voids are induced. The
peak shifts towards the right.

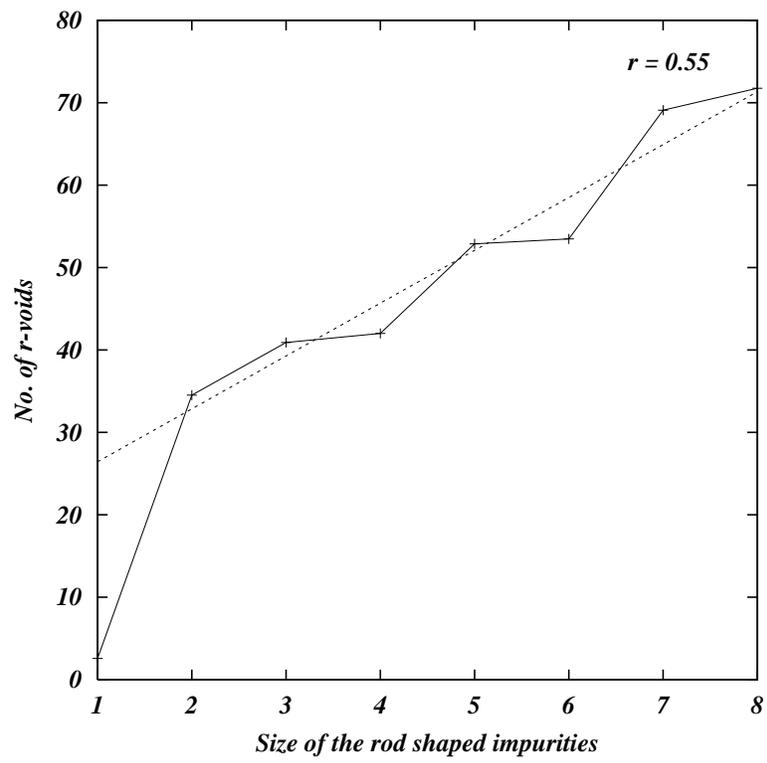
\item[{\bf Fig. 4}] Dependence of the number of $r$-voids
on the length of the rod-shaped impurities. The graph shows a steady
increase in the number of $r$-voids (for $r=0.55$) with $n$. The
first expected large jump in the number of voids because of the shape
anisotropy is seen clearly when the configuration consists of
molecules in the shape of circular disks and rod$_2$.

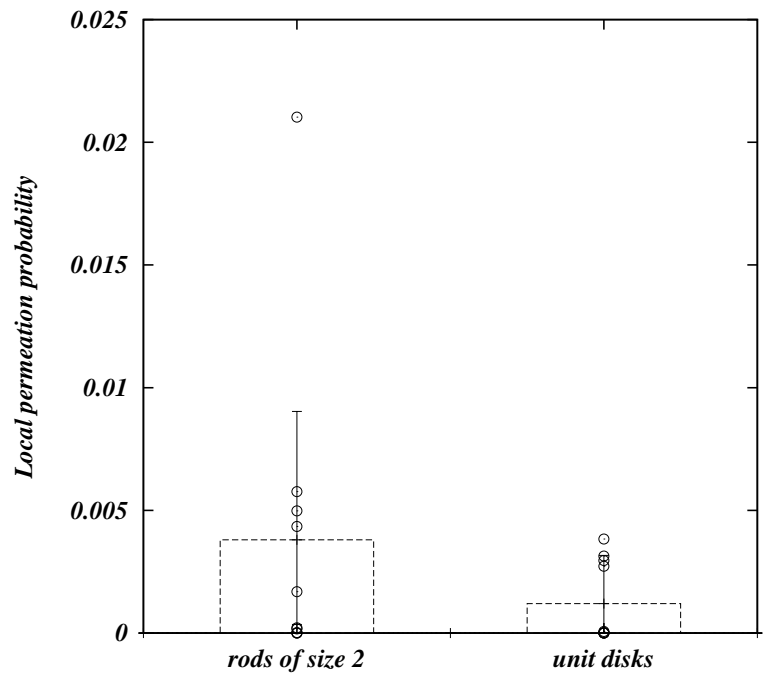
\item[{\bf Fig. 5}] Local permeation probability in a doped model
  system. The points show the {\it local permeation probability}
around ten randomly chosen unit disks and ten randomly chosen
rod$_2$s. Further, as a guide line, averages are shown by the heights
of the boxes, clearly indicating significantly more permeation in the
neighborhood of rod$_2$.
\end{description}
\end{document}